\documentclass[prb,twocolumn,showpacs]{revtex4-1}
\usepackage[T1]{fontenc}
\usepackage{verbatim}
\usepackage{graphicx}
\usepackage{graphics}
\usepackage{upgreek}
\usepackage{array}
\usepackage{mathrsfs}
\usepackage{amsmath}
\usepackage{pstricks}
\usepackage{dsfont}
\usepackage{amssymb}
\usepackage{lipsum}
\usepackage[squaren,Gray,cdot]{SIunits}
\newcommand\ii{\mathbf{i}}
\newcommand\cor[1]{\langle #1 \rangle}

\newcommand\CP{\mathcal{C}}

\newcommand\iinfty{\displaystyle\int_{-\infty}^{\infty}\,}
\newcommand\el{\textrm{el}}

\newcommand\sgn{\text{sgn}}

\begin{document}	

\title{Finite frequency noise in a quantum point contact between helical edge states}
\author{J.-R. Souquet}
\author{P. Simon}
\affiliation{Laboratoire de Physique des Solides, Universit\a'e Paris-Sud, 91405 Orsay, France}
 \date{\today}
 
\begin{abstract}
We propose and analyze the non-equilibrium finite frequency current-current correlations as a mean to characterize the helical nature of the edge states in a quantum spin hall geometry.
We show that the finite frequency noise  enables to  unambiguously discriminate between the one-particle and the two-particles processes occurring in the helical liquid
for both tunneling or weak-backscattering regimes.
\end{abstract}

\maketitle
{\em Introduction} Topological insulators have been generating a huge interest in the condensed matter community due to their exotic properties. They are electronic materials characterized by a bulk band gap like a normal insulator, but with protected conducting edge states.\cite{Hasan10,Qi11} 
Following earlier theoretical predictions,\cite{Bernevig06} the 2D topological insulating state have been realized in HgTe quantum well with an inverted band structure.\cite{Konig07} By increasing the thickness of the HgTe quantum well beyond a critical value, a standard insulator is transformed into a 
quantum spin Hall insulator (QSHI) with conducting edge channels that have been probed by non-local transport measurements.\cite{Roth09}
Due to the strong spin-orbit coupling, the electronic motion in these edge states is locked with its spin: electrons  with opposite spin on one edge propagate into opposite directions.\cite{Kane05b} These edge states are protected by time reversal symmetry against single-particle elastic backscattering. In the presence of electron interactions they form a new class of 1D liquid called helical Luttinger liquid (HLL).\cite{Xu06,Wu06,KT} However, inelastic single-particle scattering\cite{Budich12,Schmidt12} and two-particles  backscattering processes\cite{Xu06,Wu06,Strom10,Lezmy12,Crepin12} can significantly modify transport properties of the HLL.

\begin{figure}
\vspace{-40pt}
\begin{pspicture}(0,0)(7,7)
\rput[bl](-.5,0){\includegraphics[width=0.46\textwidth]{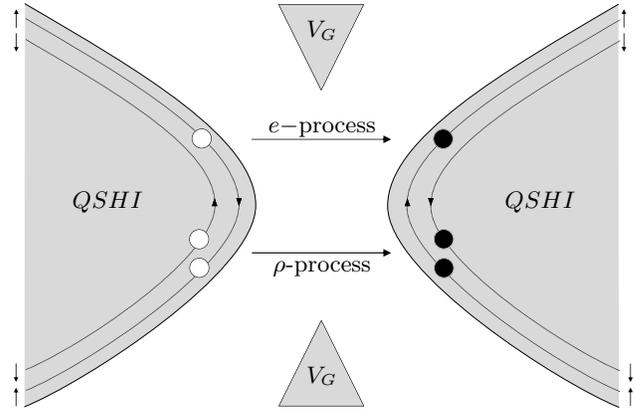}}
\rput[b](.75,2.6){$QSHI$}
\rput[b](6.5,2.6){$QSHI$}
\rput[b](3.6,4.9){$V_G$}
\rput[b](3.6,.3){$V_G$}
\rput[b](3.6,3.6){$e-$process}
\rput[b](3.6,1.75){$\rho$-process}
\end{pspicture}		
\caption{\label{sch} Sketch of the QPC geometry under consideration: In the tunneling regime, the QPC is pinched off and separates the two helical liquids}
\end{figure}

A convenient way to probe the transport properties of edge states
is to use a quantum point contact (QPC) between two  quantum spin hall insulators  as depicted in Fig. \ref{sch}. Near-equilibrium transport properties for  the HLL  have been  studied in details  in this four terminal geometry.\cite{Hou09,Strom09,KT,Schmidt11} 
Given the value of the gate voltage $V_G$, the QPC can be considered as a perfectly transmitting (the weak backscattering regime) or a perfectly reflecting (the tunneling regime) barrier for the edge states. These two limits are dual of each other. 
Because of the helical nature of this 1D liquid, it has been nicely shown in Ref. [\onlinecite{Hou09}] that this problem can be mapped to the well-studied problem of a weak link in a {\it spinful} Luttinger liquid\cite{Kane92a,Kane92b} with the Luttinger parameter in the  charge sector $g_\rho$ being the inverse of the  Luttinger parameter in the spin sector $g_\sigma$ (we therefore define in what follows $g=g_\rho=1/g_\sigma$). Despite the existence of this mapping,  
both the weak-backscattering limit (corresponding to an open QPC) and the weak tunneling limit (corresponding to an pinched-off QPC) are stable in perturbation theory for $g\in [1/2,2]$ which implies the existence of an intermediate zero temperature fixed point\cite{KT}, as opposed to the behavior of an impurity in an ordinary Luttinger liquid. Furthermore, in addition to single-particle scattering terms which usually dominate transport properties, it has been shown in [\onlinecite{KT}] that two-particle scattering terms can play an important role in this QPC geometry.

\begin{figure}[h!]
\vspace{-1.7cm}
\hspace{-10pt}
\begin{pspicture}(0,0)(7,7)
\rput[bl](-.5,0){\includegraphics[width=0.46\textwidth]{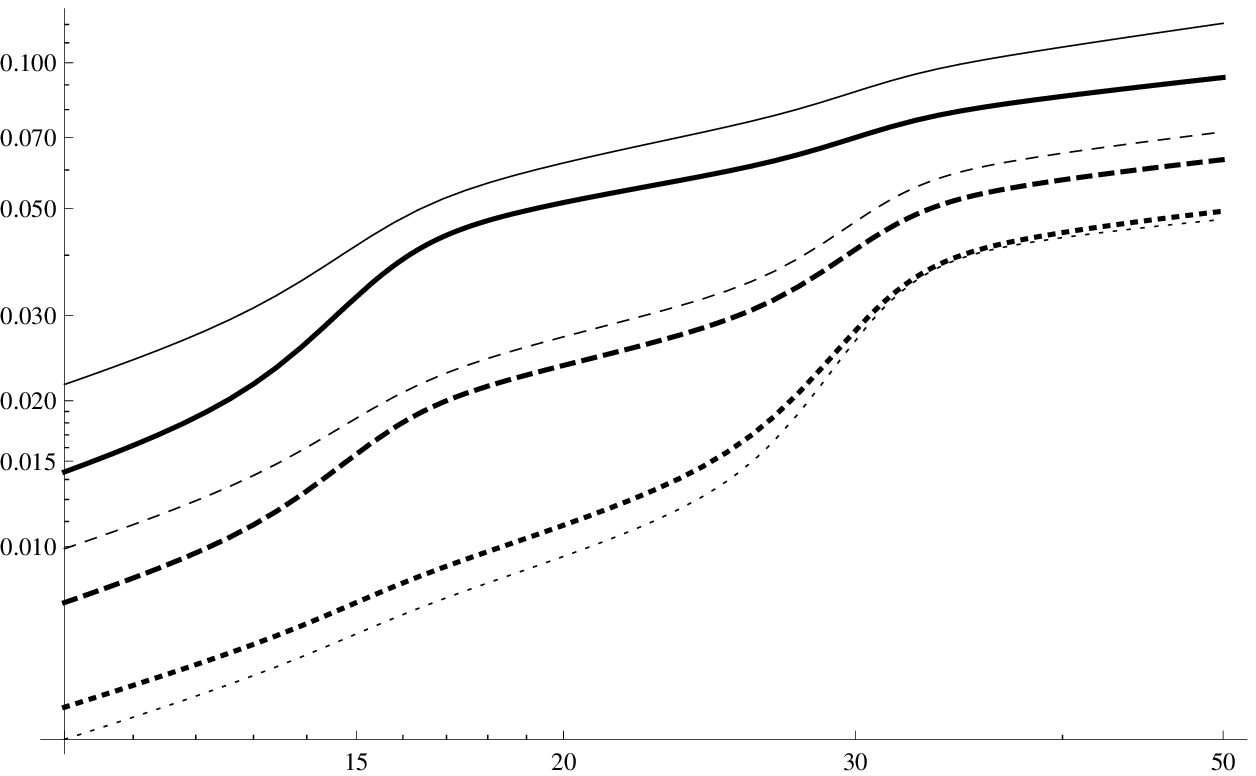}}
\rput[b](1.8517,.247){\psset{linestyle=dashed}\psline[linewidth=.5pt]{-}(0,3.8)}
\rput[b](5.135,.247){\psset{linestyle=dashed}\psline[linewidth=.5pt]{-}(0,4.4)}
\rput[b](7.4,.35){$\beta eV$}
\rput[b](2.75,.65){$\rho$-process}
\rput[b](2.75,.35){singularity}
\rput[b](6,.65){$e$-process}
\rput[b](6,.35){singularity}
\rput[b](1.05,4.5){$\dfrac{\partial_{V} S_{T}(g,eV)}{\beta eS_{T}(g,0)}$}
\end{pspicture}		
\caption{\label{St2} Ratio between the noise derivative at finite frequency and the noise at zero bias and finite frequency as a function of $\beta eV$  for different values of g and different values of  $\tau=(t_\rho/t_e)^2$. This plot is done for $\beta\omega=30$, $g=1.6$ (thick lines)and for $g=1.7$ (thin lines), for $\tau=1$ (full lines), $\tau=1/5$ (dashed lines) and $\tau=1/50$ (dotted lines).}
\end{figure}

\begin{figure}[h!]
\vspace{-1.5cm}
\hspace{-10pt}
\begin{pspicture}(0,0)(7,7)
\rput[bl](-.5,0){\includegraphics[width=0.46\textwidth]{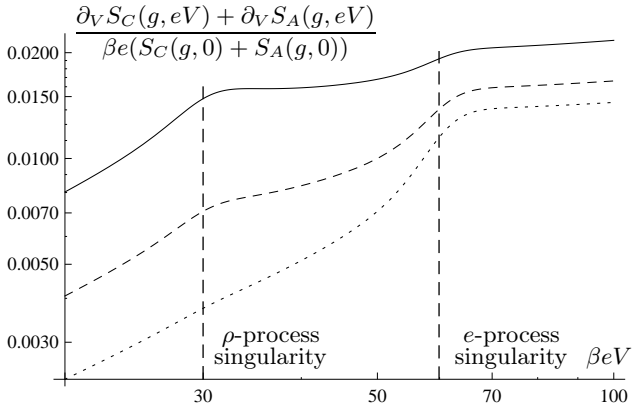}}
\rput[b](2.1,.36){\psset{linestyle=dashed}\psline[linewidth=.5pt]{-}(0,3.9)}
\rput[b](5.24,.36){\psset{linestyle=dashed}\psline[linewidth=.5pt]{-}(0,4.37)}
\rput[b](7.5,.5){$\beta eV$}
\rput[b](3,.8){$\rho$-process}
\rput[b](3,.5){singularity}
\rput[b](6.2,.8){$e$-process}
\rput[b](6.2,.5){singularity}\rput[b](2.4,4.6){$ \dfrac{\partial_{V} S_{C}(g,eV)+\partial_VS_{A}(g,eV)}{\beta e(S_{C}(g,0)+S_{A}(g,0))}$}
\end{pspicture}		
\caption{\label{ST} Ratio between the sum of the $S_A$ and the $S_C$ noise derivative at finite frequency and their value at zero bias and at finite frequency as a function of $\beta eV$  for different values of g and different ratio of $\tau=(t_\rho/t_e)^2$. This plot is done for $\beta\omega=30$, $g=0.6$ and for $\tau=1$ (full line), $\tau=1/5$ (dashed line) and $\tau=1/50$ (dotted line).}
\end{figure}

As we emphasized above, previous studies of transport properties mainly focused on the conductance and the zero-frequency noise. 
However, thanks to progress in on-chip detection of high frequency electronic properties, finite frequency (FF) transport properties are now accessible experimentally.\cite{Basset10,Basset12}
FF noise allows an access to the real time electron dynamics of the excitations in the system and may therefore constitute an appropriate experimental tool to discriminate
between one-particle and two-particle scattering processes \cite{Chamon95, Sassetti1}. The purpose of this work is therefore to present theoretical predictions  for the FF noise at finite temperature and in the perturbative regime for the quantum spin Hall bar geometry as depicted in Fig. \ref{sch}. At finite frequency $\omega$, the detection of a current fluctuation by an external circuit requires the emission of a photon of energy $\hbar\omega$. At zero temperature and when electrons do not interact, this process is forbidden as long as $(2)eV<\hbar\omega$ for one-(two)-electron processes, and when this process is allowed, singularities are expected in the finite frequency noise. We find that the single and two-particle processes can be clearly distinguished in the FF noise (more exactly in its derivative) in both the tunneling and weak backscattering regime as can be seen in  Fig. \ref{St2} and Fig. \ref{ST}.

{\em Model and notations} 
We follow Teo and Kane \cite{KT} and directly map the problem on the one of a tunneling electron between spinful Luttinger liquids. The Hamiltonian of the edge state reads
\begin{equation}
H=\frac{u}{4\pi}\iinfty dx \sum_{a=\sigma,\rho}g_a(\partial_x\phi_a)^2+\frac{1}{g_a}(\partial_x\theta_a)^2,
\label{eq:}
\end{equation}
where $u$ is the effective Fermi velocity, $\phi_a$ and $\nabla\theta_a$ are two conjugate fields that describe the liquids and such that $\left[\theta_a,\phi_b\right] =2\ii\pi\delta_{a,b}\theta(x-y)$ with $\delta_{a,b}$ the Kronecker symbol and $\theta(x)$ the Heaviside function. We have set $\hbar=1$ throughout the paper.

The action of the QPC is modeled by a local potential that respects time reversal symmetry. It has different relevant terms depending on the strength of Coulomb interactions and the strength of the gate voltage of the QPC. If we focus on spin conserved interactions, four different fixed points are found.\cite{KT}
Following the notation of [\onlinecite{KT}], we denote $V_{CC}$ the perfectly transmitting potential for both charge and spin sectors, $V_{II}$ the perfectly reflecting potential for both sectors, and finally
$V_{CI}$ (resp. $V_{IC}$) the potential when the QPC is perfectly transmitting (resp. reflecting) for the charge and reflecting (resp. transmitting) for the spin.

The ${CC}$ and ${CI}$ cases correspond to the weak backscattering regime while the two others to the tunnel regime. For all cases, single-electronic processes are labeled in the following by a $e$ subscript. In the $V_{CC}$ and the $V_{II}$ cases, two-particles processes have to be also taken into account and are labeled by the $g$ subscripts.\cite{KT} The expression for each of these interacting potential, their renormalization group flow as well as their domain of validity  is detailed in [\onlinecite{KT}].

\begin{figure}[h!]
\vspace{-1.5cm}
\hspace{-10pt}
\begin{pspicture}(0,0)(7,7)
\rput[bl](-.5,0){\includegraphics[width=0.46\textwidth]{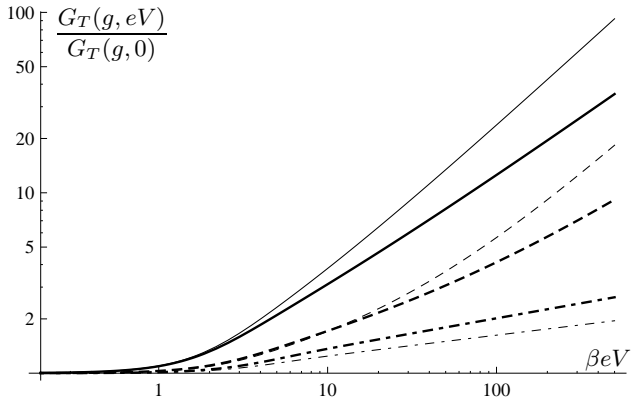}}
\rput[b](7.5,.4){$\beta eV$}
\rput[b](.9,4.5){$\dfrac{G_{T}(g,eV)}{G_{T}(g,0)}$}
\end{pspicture}		
\caption{\label{Gt05} Ratio between the differential conductance and its value at zero bias as a function of $\beta eV$ at fixed $\beta$ for different values of g and different ratio of $\tau=(t_\rho/t_e)^2$. This plot is done for $g=1.5$ (thick lines)and for $g=1.4$ (thin lines), for $\tau=1$ (full lines), $\tau=1/5$ (dashed lines) and $\tau=0$ (dotdashed lines).}
\end{figure}

Last we impose a d.c. bias to the edge fields. In both conducting limits, the QPC behaves as a capacitance and we assume that the voltage drop takes place at its edges. This voltage drop can be taken into account by performing a gauge transformation on the scattering potential. In the tunneling limit this achieved by changing $\theta_\rho(0,t)\to\theta_\rho(0,t)-eVt$ and in the transparent regime by $\phi_\rho(0,t)\to\phi_\rho(0,t)-e^*Vt$ with $e^*=ge$ the effective charge. We deal with the out-of-equilibrium situation by using the Keldysh path integral formalism.\cite{Keldysh} The Green functions of the LL are written:
\begin{equation}
\begin{array}{rl}
\CP_{a}^{^{\pm\pm}}(x,t)=&\frac{1}{2g_a}\ln\left(\frac{\exp(\tfrac{2\pi |x|}{\beta u})}{2\left|\sinh^2(\tfrac{\pi t}{\beta})-\sinh^2(\tfrac{\pi x}{\beta u})\right|}\right)\\
&\pm\frac{\ii\pi}{2g_a}\Theta\left(|t|-\dfrac{|x|}{u}\right),\\
\end{array}
\label{eq:cpp}
\end{equation}
\begin{equation}
\begin{array}{rl}
\CP_{a}^{^{\pm\mp}}(x,t)	=&\frac{1}{2g_a}\ln\left(\frac{\exp(\tfrac{2\pi |x|}{\beta u})}{2\left|\sinh^2(\tfrac{\pi t}{\beta})-\sinh^2(\tfrac{\pi x}{\beta u})\right|}\right)\\
&\pm\frac{\ii\pi}{2g_a}\sgn(t)\Theta\left(|t|-\dfrac{|x|}{u}\right).
\end{array}
\label{eq:cpm}
\end{equation}

where the $\eta=\pm$ superscripts labels respectively the upper and the lower branch of the Keldysh contour.\cite{Keldysh} 
{\em Results} We now compute for both the tunneling and weak backscattering regime, the conductance and the noise at finite temperature, finite voltage and finite frequency for the QPC. 

i) {\it The tunneling regime (II)} This phase is stable for $g\in[1/2;2]$. Using the same notations as in [\onlinecite{KT}], the potential  $V_{II}$ reads:
\begin{equation}
V_{II}=t_e\cos(\tilde\theta_\rho+\eta_\rho)\cos\tilde\theta_\sigma+t_\rho\cos 2\tilde{\theta}_{\rho}+t_\sigma\cos 2\tilde \theta_\sigma,
\label{eq:vii}
\end{equation}
with $\tilde\theta_a=(\phi_a^{\textrm{right}}-\phi_a^{\textrm{left}})/2$. The term in $t_e$ represents the spin-conserved tunneling of a single electron from one side to the other and the phase $\cos\tilde\theta_\sigma$ is fixed by time-reversal symmetry. The phase $\eta_\rho$ is arbitrary. $t_\sigma$ and $t_\rho$ are associated with two-electron processes: $t_\sigma$ represents an exchange of an electron on each side of the barrier and $t_\rho$ the tunneling of two electrons of opposite spin on the same liquid to the other side of the barrier. From this analysis, it is obvious that only $\rho-$ and $e-$ processes will contribute to the tunneling current. We define the tunneling current operator as the following quantity:
\begin{equation}
j^{\eta}_{T}(t)=-\frac{e}{\hbar}\displaystyle\int d\tau \displaystyle \frac{\updelta V_{\alpha\alpha'}[\theta^\eta_\rho(\tau)+e^*V(\tau)]}{\updelta \theta^\eta_\rho(t)}.
\label{eq:jeta}
\end{equation}
The tunneling current $I_T$ is then defined as the average value of the previous quantity:
\begin{equation}
I_T=e\left(2t_\rho^2\Gamma_{1/g}(2eV)+t_e^2\Gamma_{\tilde g}(eV)\right),
\label{eq:it}
\end{equation} 
with $\tilde g=(g+g^{-1})/4$ and:
\begin{equation}
\Gamma_g(E)=\frac{2\pi\beta}{\Gamma(4g)}\left(\tfrac{\pi\alpha}{\beta u}\right)^{4g} \left|\Gamma(2g-\ii\tfrac{\beta E}{2\pi})\right|^2\sinh\left(\tfrac{\beta E}{2}\right),
\label{eq:LNK}
\end{equation}
where $\alpha$ is the interatomic distance and $(\pi\alpha/\beta u)$ a high energy cut-off. We also introduce a {\it phenomenological} parameter $\tau=(t_\rho/t_e)^2$ factor that compares the relative probability of each process. The two $\Gamma$ terms in Eq. (\ref{eq:it}) can be interpreted as the  probability for a charge or a pair of charge to tunnel through the QPC. These two processes have significantly different power laws with respect to the bias $V$. Single electron processes (associated with $t_e$) follow a $V^{2\tilde g-1}$ power law, whereas the two-electrons processes ( the $t_\rho$ term) follow a $V^{4/g-1}$.
These different scaling behaviors are manifest in the differential conductance $G(g,V)$, which  is plotted as a function of $\beta eV$ in Fig.\ref{Gt05}. In this figure two regimes clearly emerge: the high temperature regime ($\beta eV\ll1$) for which the current is temperature driven and linear, and a low temperature regime $\beta eV\gg1$ for which the current is driven by Coulomb interactions. Depending on the values of the parameter $\tau$ and of the voltage, the leading mechanism will either be $e-$ processes (dotdashed lines) or $\rho-$processes (full lines). The distinction can be made only by a precise analysis of power law exponents which may be experimentally difficult. Moreover, the approximations used to obtain these results bound us to voltage much smaller than the Fermi energy. Besides, at frequencies larger than the level spacing in the QPC, non-local transport features need to be taken into account\cite{tztl2,Sassetti2} which may affect power laws and make the distinction between single- and two-particles processes even more laborious. 

To go beyond this difficulty we compute the FF tunneling noise, which should be zero as long as $eV<\hbar\omega$ ($2eV<\hbar\omega$) for $e-$process ($\rho-$process)\cite{Glattli}. Therefore if the latter is to be of any importance, a singularity in the noise derivative should appear at $2eV=\hbar\omega$. In addition, by arbitrary fixing the finite frequency (much larger than temperature to be in the quantum regime), we can probe high voltage regimes where $\rho$-processes are likely to be of greater significance. The tunneling noise reads $S_T=\cor{j_T(t)j_T(0)}-\cor{j_T(0)}^2$. Note that we are only considering {\it local} current-current correlations in this work. We can neglect the second term which is of fourth order in $t_a$ to obtain:
\begin{equation}
\begin{array}{rl}
S_T(\omega)&=4e\displaystyle\sum_{\eta}\coth(\beta(\omega+\eta eV))
t_\rho^2\Gamma_{1/g}(\omega+2\eta eV)\\
&+8e\displaystyle\sum_{\eta}\coth(\beta(\omega+2\eta eV)){4}t_e^2\Gamma_{\tilde g}(\omega+\eta eV).
\end{array}
\label{eq:ST}
\end{equation}
At first order in perturbation theory, there is no correlation between single and two-particle scattering modes. The noise can be considered as the sum of two independent sources, one due to $e$- and another one due to $\rho-$ processes. Despite the out-of-equilibrium situation, we notice that there is a  fluctuation-dissipation relation between noise and current for each process as this has already been reported in other systems\cite{Safi07}. 
In both cases, a singularity appears respectively at $eV=\omega$ and $2eV=\omega$ whose size depends on the strength of each scattering processes. This singularity is apparent at $g=2$ for $\rho-$processes and
at $g=1$ for $e-$processes. As it is rounded off by temperature and interactions, this singularity appears more clearly on the noise derivative as can be seen on Fig. \ref{St2} (note that the derivative of the FF noise was the quantity directly extracted experimentally in Ref. [\onlinecite{Basset12,Glattli}]). 
For $e-$process, the noise derivative power law is always smaller than one, meaning the singularity cannot be completely blurred out by interactions. Therefore if $\rho-$processes are negligeable, experimental measurements should recover results similar to the one from Ref[\onlinecite{Glattli}]. For $\rho-$processes, the noise can follow a $V^{4/g-1}$ power law. Its singularity compared to the competitive process singularity will be burred if $\log(\tau)\gg 4(\tilde g-g^{-1})\log(2\hbar \omega)$. However, the $\rho-$process singularity appear more patently when higher derivatives of the noise are plotted. 
ii) {\it The weak backscattering regime (CC)}
 The CC phase is also stable for $g\in[1/2;2]$ and the interacting Hamiltonian due to the QPC reads now:
\begin{equation}
V_{CC}=v_e\cos(\tilde\theta_\rho+\eta_\rho)\cos\tilde\theta_\sigma+v_\rho\cos 2\tilde{\theta}_{\rho}+v_\sigma\cos 2 \tilde\theta_\sigma,
\label{eq:}
\end{equation}
where $t_e$ represents the elementary tunneling of a charge at the QPC ($e-$process). The $v_\rho$ term involves the backscattering of a pair of electron ($\rho$-process) of opposite spin and  $v_\sigma$ embodies the tunneling of a unit of spin from the right to the left moving channels and a tunneling of charge $2e^*$ between the top and the bottom edges. Again, the two processes can contribute to the backscattering current are $v_e$ and $v_\rho$.
In the weak backscattering case, the current can be written as the contribution of the bare current $I_0=2V/R_K$ with $R_K$ the von Klitzling resistance and the weak backscattering current $I_B$: $I=I_0-I_B$. The backscattering current operator can be defined as:\cite{tztl}
\begin{equation}
j_B^\eta=-\frac{e}{\hbar }\lim_{J\to0}\displaystyle\int d\tau \frac{\updelta V_{CC}[\theta^\eta_\rho(\tau)+e^*V\tau+A_J(\tau)]}{\updelta J(t)}
,\label{eq:jb}
\end{equation}
with:
\begin{equation}
A_J(\tau)=\displaystyle \int d\tau'\left(\CP^K_{\rho}(\tau-\tau')+\eta\ii\CP_{\rho}^A(\tau-\tau')\right)J(\tau'),
\label{eq:}
\end{equation}
where $C^K$ and $C^A$ are respectively the Keldysh and the advanced Green functions. The tunneling current is then defined as the average value of Eq. \eqref{eq:jb} and can be computed following [\onlinecite{tztl}]: 
\begin{equation}
I_B(V)=\frac{1}{2}\displaystyle\sum_{\eta}\iinfty dt' \partial_tC^R_{\el}(t-t')\cor{j^\eta_B}(t'),
\label{eq:}
\end{equation}
where $C^R_{\el}(t)=2gR_K^{-1}\theta(t)$. At first order in perturbation, the current is time independent and $I_B=\cor{j_B}(0)$  writes as
\begin{equation}
I_B(V)=2e\left(v_\rho^2\Gamma_g(2e^*V)+\frac{v_e^2}{2}\Gamma_{\tilde g}(e^*V)\right).
\label{eq:}
\end{equation}

This is worth emphasizing that the backscattering current can be directly recovered  from the tunneling case (II)  by performing the transformation $g\to1/g$. The $e$-processes scattering are invariant under such transformation and thus remain unchanged compared to the tunneling case. However the underlying physics is different since the spin and the charge have exchanged their role in the enhancing of the current. Besides, since the domain of stability of the (II) and (CC) phase is symmetric under this transformation, Fig.\ref{Gt05} represents as well the differential back scattering conductance for $g=1$ and $g=1.4$.   
The main difference between the two conducting limits lies in the second cumulant. As shown in [\onlinecite{tztl}], it is the contribution of three different sources: $S_0$, the bare current-current correlations, $S_A$, the backscattering current-current correlation or shot noise, and $S_C$, which embodies retarded effects of the backscattering current on noise. They can be compactly expressed in the case of a Luttinger liquid as:
\begin{eqnarray}
&S_0(\omega)=&\displaystyle (e\omega)^2\CP_\rho^K(\omega),\\
&S_A(\omega)=&ev_\rho^2\left(\omega\CP^R(\omega)\right)^2\sum_{\eta} \coth\left(\beta(\omega+2\eta e^*V)\right) \label{eq:SA}\\
&&\times \Gamma_{g}(\omega+2\eta e^*V)+\dfrac{ev_e^2}{4}\left(\omega\CP^R(\omega)\right)^2 \nonumber\\
&&\times\sum_{\eta} \coth\left(\beta(\omega+\eta e^*V)\right)\Gamma_{\tilde g}(\omega+\eta e^*V),\nonumber\\
&S_C(\omega)=&-e\omega^2\CP_{\rho}^K(\omega)\CP_{\rho}^R(\omega)\displaystyle\sum_{\eta\eta'}\eta' I_B(\eta'\omega+\eta geV).
\end{eqnarray} 
Depending on the strength of interactions, bias and frequency, the excess noise is mainly due to shot noise.\cite{tztl} Again, singularities appear respectively at $e^*V=\omega/2$ and $e^*V=\omega$ as can be observed on Fig.\ref{ST}.

iii) {\it The (IC) and (CI) cases}
We finally briefly comment on the (IC) and (CI) cases. The (IC) fixed point is stable for $g<1/2$. In this regime, the scattering potential only contains  single-electronic terms.\cite{KT} The tunneling current and the shot noise can be thus obtained directly from the II case by  only keeping $e-$process terms and by setting $\tilde g=1/4g$. Similarly the (CI) case (which is stable for $g>2$) is obtained in a similar way where one has to set $\tilde g=g/4$ and change $e$ into its effective charge $e^*=ge$.

{\em Summary} To summarize, we have shown in this work how the FF noise enables a clear distinction between one-particle and two-particles scattering processes. 
For both the tunneling and weak-backscattering regimes,
these processes manifest through a singularity at either $eV=\hbar\omega$ (one particle) or $eV=\hbar \omega/2$ (two-particles) as can be for example seen in Figs.\ref{St2} and \ref{ST}.

{\it Notes added} During the final course of preparation of this
manuscript, we notice Ref.[\onlinecite{Chung}] which has  partial overlap with the present work.
\bibliographystyle{apsrev4-1}
\bibliography{Helical}

\begin{thebibliography}{10}%
\makeatletter
\providecommand \@ifxundefined [1]{%
 \ifx #1\undefined \expandafter \@firstoftwo
 \else \expandafter \@secondoftwo
\fi
}%
\providecommand \@ifnum [1]{%
 \ifnum #1\expandafter \@firstoftwo
 \else \expandafter \@secondoftwo
\fi
}%
\providecommand \enquote [1]{``#1''}%
\providecommand \bibnamefont  [1]{#1}%
\providecommand \bibfnamefont [1]{#1}%
\providecommand \citenamefont [1]{#1}%
\providecommand\href[0]{\@sanitize\@href}%
\providecommand\@href[1]{\endgroup\@@startlink{#1}\endgroup\@@href}%
\providecommand\@@href[1]{#1\@@endlink}%
\providecommand \@sanitize [0]{\begingroup\catcode`\&12\catcode`\#12\relax}%
\@ifxundefined \pdfoutput {\@firstoftwo}{%
 \@ifnum{\z@=\pdfoutput}{\@firstoftwo}{\@secondoftwo}%
}{%
 \providecommand\@@startlink[1]{\leavevmode\special{html:<a href="#1">}}%
 \providecommand\@@endlink[0]{\special{html:</a>}}%
}{%
 \providecommand\@@startlink[1]{%
  \leavevmode
  \pdfstartlink
   attr{/Border[0 0 1 ]/H/I/C[0 1 1]}%
   user{/Subtype/Link/A<</Type/Action/S/URI/URI(#1)>>}%
  \relax
 }%
 \providecommand\@@endlink[0]{\pdfendlink}%
}%
\providecommand \url  [0]{\begingroup\@sanitize \@url }%
\providecommand \@url [1]{\endgroup\@href {#1}{\urlprefix}}%
\providecommand \urlprefix [0]{URL }%
\providecommand \Eprint[0]{\href }%
\@ifxundefined \urlstyle {%
  \providecommand \doi [1]{doi:\discretionary{}{}{}#1}%
}{%
  \providecommand \doi [0]{doi:\discretionary{}{}{}\begingroup
  \urlstyle{rm}\Url }%
}%
\providecommand \doibase [0]{http://dx.doi.org/}%
\providecommand \Doi[1]{\href{\doibase#1}}%
\providecommand \bibAnnote [3]{%
  \BibitemShut{#1}%
  \begin{quotation}\noindent
    \textsc{Key:}\ #2\\\textsc{Annotation:}\ #3%
  \end{quotation}%
}%
\providecommand \bibAnnoteFile [2]{%
  \IfFileExists{#2}{\bibAnnote {#1} {#2} {\input{#2}}}{}%
}%
\providecommand \typeout [0]{\immediate \write \m@ne }%
\providecommand \selectlanguage [0]{\@gobble}%
\providecommand \bibinfo [0]{\@secondoftwo}%
\providecommand \bibfield [0]{\@secondoftwo}%
\providecommand \translation [1]{[#1]}%
\providecommand \BibitemOpen[0]{}%
\providecommand \bibitemStop [0]{}%
\providecommand \bibitemNoStop [0]{.\EOS\space}%
\providecommand \EOS [0]{\spacefactor3000\relax}%
\providecommand \BibitemShut [1]{\csname bibitem#1\endcsname}%
\bibitem{Hasan10}%
  \BibitemOpen
  \bibfield{author}{%
  \bibinfo {author} {\bibfnamefont{M.~Z.}\ \bibnamefont{Hasan}}\ and\ \bibinfo
  {author} {\bibfnamefont{C.~L.}\ \bibnamefont{Kane}},\ }%
  \bibfield{journal}{%
  \Doi{10.1103/RevModPhys.82.3045}{\bibinfo {journal} {Rev. Mod. Phys.}}\ }%
  \textbf{\bibinfo {volume} {82}},\ \bibinfo {pages} {3045} (\bibinfo {month}
  {Nov}\ \bibinfo {year} {2010}),\
  \url{http://link.aps.org/doi/10.1103/RevModPhys.82.3045}%
  \bibAnnoteFile{NoStop}{Hasan10}%
\bibitem{Qi11}%
  \BibitemOpen
  \bibfield{author}{%
  \bibinfo {author} {\bibfnamefont{X.-L.}\ \bibnamefont{Qi}}\ and\ \bibinfo
  {author} {\bibfnamefont{S.-C.}\ \bibnamefont{Zhang}},\ }%
  \bibfield{journal}{%
  \Doi{10.1103/RevModPhys.83.1057}{\bibinfo {journal} {Rev. Mod. Phys.}}\ }%
  \textbf{\bibinfo {volume} {83}},\ \bibinfo {pages} {1057} (\bibinfo {month}
  {Oct}\ \bibinfo {year} {2011}),\
  \url{http://link.aps.org/doi/10.1103/RevModPhys.83.1057}%
  \bibAnnoteFile{NoStop}{Qi11}%
\bibitem{Bernevig06}%
  \BibitemOpen
  \bibfield{author}{%
  \bibinfo {author} {\bibfnamefont{B.~A.}\ \bibnamefont{Bernevig}}, \bibinfo
  {author} {\bibfnamefont{T.~L.}\ \bibnamefont{Hughes}},\ and\ \bibinfo
  {author} {\bibfnamefont{S.~C.}\ \bibnamefont{Zhang}},\ }%
  \bibfield{journal}{%
  \bibinfo {journal} {Science}\ }%
  \textbf{\bibinfo {volume} {314}},\ \bibinfo {pages} {1757} (\bibinfo {year}
  {2006})%
  \bibAnnoteFile{NoStop}{Bernevig06}%
\bibitem{Konig07}%
  \BibitemOpen
  \bibfield{author}{%
  \bibinfo {author} {\bibfnamefont{M.}~\bibnamefont{Koenig}}\ and\ \bibinfo
  {author} {\bibnamefont{et~al.}},\ }%
  \bibfield{journal}{%
  \bibinfo {journal} {Science}\ }%
  \textbf{\bibinfo {volume} {318}},\ \bibinfo {pages} {766} (\bibinfo {year}
  {2007})%
  \bibAnnoteFile{NoStop}{Konig07}%
\bibitem{Roth09}%
  \BibitemOpen
  \bibfield{author}{%
  \bibinfo {author} {\bibfnamefont{A.}~\bibnamefont{Roth}}\ and\ \bibinfo
  {author} {\bibnamefont{et~al.}},\ }%
  \bibfield{journal}{%
  \bibinfo {journal} {Science}\ }%
  \textbf{\bibinfo {volume} {325}},\ \bibinfo {pages} {294} (\bibinfo {year}
  {2009})%
  \bibAnnoteFile{NoStop}{Roth09}%
\bibitem{Kane05b}%
  \BibitemOpen
  \bibfield{author}{%
  \bibinfo {author} {\bibfnamefont{C.~L.}\ \bibnamefont{Kane}}\ and\ \bibinfo
  {author} {\bibfnamefont{E.~J.}\ \bibnamefont{Mele}},\ }%
  \bibfield{journal}{%
  \Doi{10.1103/PhysRevLett.95.226801}{\bibinfo {journal} {Phys. Rev. Lett.}}\
  }%
  \textbf{\bibinfo {volume} {95}},\ \bibinfo {pages} {226801} (\bibinfo {month}
  {Nov}\ \bibinfo {year} {2005}),\
  \url{http://link.aps.org/doi/10.1103/PhysRevLett.95.226801}%
  \bibAnnoteFile{NoStop}{Kane05b}%
\bibitem{Xu06}%
  \BibitemOpen
  \bibfield{author}{%
  \bibinfo {author} {\bibfnamefont{C.}~\bibnamefont{Xu}}\ and\ \bibinfo
  {author} {\bibfnamefont{J.~E.}\ \bibnamefont{Moore}},\ }%
  \bibfield{journal}{%
  \Doi{10.1103/PhysRevB.73.045322}{\bibinfo {journal} {Phys. Rev. B}}\ }%
  \textbf{\bibinfo {volume} {73}},\ \bibinfo {pages} {045322} (\bibinfo {month}
  {Jan}\ \bibinfo {year} {2006}),\
  \url{http://link.aps.org/doi/10.1103/PhysRevB.73.045322}%
  \bibAnnoteFile{NoStop}{Xu06}%
\bibitem{Wu06}%
  \BibitemOpen
  \bibfield{author}{%
  \bibinfo {author} {\bibfnamefont{C.}~\bibnamefont{Wu}}, \bibinfo {author}
  {\bibfnamefont{B.~A.}\ \bibnamefont{Bernevig}},\ and\ \bibinfo {author}
  {\bibfnamefont{S.-C.}\ \bibnamefont{Zhang}},\ }%
  \bibfield{journal}{%
  \Doi{10.1103/PhysRevLett.96.106401}{\bibinfo {journal} {Phys. Rev. Lett.}}\
  }%
  \textbf{\bibinfo {volume} {96}},\ \bibinfo {pages} {106401} (\bibinfo {month}
  {Mar}\ \bibinfo {year} {2006}),\
  \url{http://link.aps.org/doi/10.1103/PhysRevLett.96.106401}%
  \bibAnnoteFile{NoStop}{Wu06}%
\bibitem{KT}%
  \BibitemOpen
  \bibfield{author}{%
  \bibinfo {author} {\bibfnamefont{J.~C.~Y.}\ \bibnamefont{Teo}}\ and\ \bibinfo
  {author} {\bibfnamefont{C.~L.}\ \bibnamefont{Kane}},\ }%
  \bibfield{journal}{%
  \Doi{10.1103/PhysRevB.79.235321}{\bibinfo {journal} {Phys. Rev. B}}\ }%
  \textbf{\bibinfo {volume} {79}},\ \bibinfo {pages} {235321} (\bibinfo {month}
  {Jun}\ \bibinfo {year} {2009}),\
  \url{http://link.aps.org/doi/10.1103/PhysRevB.79.235321}%
  \bibAnnoteFile{NoStop}{KT}%
\bibitem{Budich12}%
  \BibitemOpen
  \bibfield{author}{%
  \bibinfo {author} {\bibfnamefont{J.~C.}\ \bibnamefont{Budich}}, \bibinfo
  {author} {\bibfnamefont{F.}~\bibnamefont{Dolcini}}, \bibinfo {author}
  {\bibfnamefont{P.}~\bibnamefont{Recher}},\ and\ \bibinfo {author}
  {\bibfnamefont{B.}~\bibnamefont{Trauzettel}},\ }%
  \bibfield{journal}{%
  \Doi{10.1103/PhysRevLett.108.086602}{\bibinfo {journal} {Phys. Rev. Lett.}}\
  }%
  \textbf{\bibinfo {volume} {108}},\ \bibinfo {pages} {086602} (\bibinfo
  {month} {Feb}\ \bibinfo {year} {2012}),\
  \url{http://link.aps.org/doi/10.1103/PhysRevLett.108.086602}%
  \bibAnnoteFile{NoStop}{Budich12}%
\bibitem{Schmidt12}%
  \BibitemOpen
  \bibfield{author}{%
  \bibinfo {author} {\bibfnamefont{T.~L.}\ \bibnamefont{Schmidt}}, \bibinfo
  {author} {\bibfnamefont{S.}~\bibnamefont{Rachel}}, \bibinfo {author}
  {\bibfnamefont{F.}~\bibnamefont{von Oppen}},\ and\ \bibinfo {author}
  {\bibfnamefont{L.~I.}\ \bibnamefont{Glazman}},\ }%
  \bibfield{journal}{%
  \Doi{10.1103/PhysRevLett.108.156402}{\bibinfo {journal} {Phys. Rev. Lett.}}\
  }%
  \textbf{\bibinfo {volume} {108}},\ \bibinfo {pages} {156402} (\bibinfo
  {month} {Apr}\ \bibinfo {year} {2012}),\
  \url{http://link.aps.org/doi/10.1103/PhysRevLett.108.156402}%
  \bibAnnoteFile{NoStop}{Schmidt12}%
\bibitem{Strom10}%
  \BibitemOpen
  \bibfield{author}{%
  \bibinfo {author} {\bibfnamefont{A.}~\bibnamefont{Str\"om}}, \bibinfo
  {author} {\bibfnamefont{H.}~\bibnamefont{Johannesson}},\ and\ \bibinfo
  {author} {\bibfnamefont{G.~I.}\ \bibnamefont{Japaridze}},\ }%
  \bibfield{journal}{%
  \Doi{10.1103/PhysRevLett.104.256804}{\bibinfo {journal} {Phys. Rev. Lett.}}\
  }%
  \textbf{\bibinfo {volume} {104}},\ \bibinfo {pages} {256804} (\bibinfo
  {month} {Jun}\ \bibinfo {year} {2010}),\
  \url{http://link.aps.org/doi/10.1103/PhysRevLett.104.256804}%
  \bibAnnoteFile{NoStop}{Strom10}%
\bibitem{Lezmy12}%
  \BibitemOpen
  \bibfield{author}{%
  \bibinfo {author} {\bibfnamefont{N.}~\bibnamefont{Lezmy}}, \bibinfo {author}
  {\bibfnamefont{Y.}~\bibnamefont{Oreg}},\ and\ \bibinfo {author}
  {\bibfnamefont{M.}~\bibnamefont{Berkooz}}}%
   (\bibinfo {year} {2012}),\
  \Eprint{http://arxiv.org/abs/1201.6197}{arXiv:1201.6197}%
  \bibAnnoteFile{NoStop}{Lezmy12}%
\bibitem{Crepin12}%
  \BibitemOpen
  \bibfield{author}{%
  \bibinfo {author} {\bibfnamefont{F.}~\bibnamefont{Cr\a'epin}}, \bibinfo
  {author} {\bibfnamefont{J.~C.}\ \bibnamefont{Budich}}, \bibinfo {author}
  {\bibfnamefont{F.}~\bibnamefont{Dolcini}}, \bibinfo {author}
  {\bibfnamefont{P.}~\bibnamefont{Recher}},\ and\ \bibinfo {author}
  {\bibfnamefont{B.}~\bibnamefont{Trauzettel}}}%
   (\bibinfo {year} {2012}),\
  \Eprint{http://arxiv.org/abs/1205.0374}{arXiv:1205.0374}%
  \bibAnnoteFile{NoStop}{Crepin12}%
\bibitem{Hou09}%
  \BibitemOpen
  \bibfield{author}{%
  \bibinfo {author} {\bibfnamefont{C.-Y.}\ \bibnamefont{Hou}}, \bibinfo
  {author} {\bibfnamefont{E.-A.}\ \bibnamefont{Kim}},\ and\ \bibinfo {author}
  {\bibfnamefont{C.}~\bibnamefont{Chamon}},\ }%
  \bibfield{journal}{%
  \Doi{10.1103/PhysRevLett.102.076602}{\bibinfo {journal} {Phys. Rev. Lett.}}\
  }%
  \textbf{\bibinfo {volume} {102}},\ \bibinfo {pages} {076602} (\bibinfo
  {month} {Feb}\ \bibinfo {year} {2009}),\
  \url{http://link.aps.org/doi/10.1103/PhysRevLett.102.076602}%
  \bibAnnoteFile{NoStop}{Hou09}%
\bibitem{Strom09}%
  \BibitemOpen
  \bibfield{author}{%
  \bibinfo {author} {\bibfnamefont{A.}~\bibnamefont{Str\"om}}\ and\ \bibinfo
  {author} {\bibfnamefont{H.}~\bibnamefont{Johannesson}},\ }%
  \bibfield{journal}{%
  \Doi{10.1103/PhysRevLett.102.096806}{\bibinfo {journal} {Phys. Rev. Lett.}}\
  }%
  \textbf{\bibinfo {volume} {102}},\ \bibinfo {pages} {096806} (\bibinfo
  {month} {Mar}\ \bibinfo {year} {2009}),\
  \url{http://link.aps.org/doi/10.1103/PhysRevLett.102.096806}%
  \bibAnnoteFile{NoStop}{Strom09}%
\bibitem{Schmidt11}%
  \BibitemOpen
  \bibfield{author}{%
  \bibinfo {author} {\bibfnamefont{T.~L.}\ \bibnamefont{Schmidt}},\ }%
  \bibfield{journal}{%
  \Doi{10.1103/PhysRevLett.107.096602}{\bibinfo {journal} {Phys. Rev. Lett.}}\
  }%
  \textbf{\bibinfo {volume} {107}},\ \bibinfo {pages} {096602} (\bibinfo
  {month} {Aug}\ \bibinfo {year} {2011}),\
  \url{http://link.aps.org/doi/10.1103/PhysRevLett.107.096602}%
  \bibAnnoteFile{NoStop}{Schmidt11}%
\bibitem{Kane92a}%
  \BibitemOpen
  \bibfield{author}{%
  \bibinfo {author} {\bibfnamefont{C.~L.}\ \bibnamefont{Kane}}\ and\ \bibinfo
  {author} {\bibfnamefont{M.~P.~A.}\ \bibnamefont{Fisher}},\ }%
  \bibfield{journal}{%
  \Doi{10.1103/PhysRevLett.68.1220}{\bibinfo {journal} {Phys. Rev. Lett.}}\ }%
  \textbf{\bibinfo {volume} {68}},\ \bibinfo {pages} {1220} (\bibinfo {month}
  {Feb}\ \bibinfo {year} {1992}),\
  \url{http://link.aps.org/doi/10.1103/PhysRevLett.68.1220}%
  \bibAnnoteFile{NoStop}{Kane92a}%
\bibitem{Kane92b}%
  \BibitemOpen
  \bibfield{author}{%
  \bibinfo {author} {\bibfnamefont{C.~L.}\ \bibnamefont{Kane}}\ and\ \bibinfo
  {author} {\bibfnamefont{M.~P.~A.}\ \bibnamefont{Fisher}},\ }%
  \bibfield{journal}{%
  \Doi{10.1103/PhysRevB.46.15233}{\bibinfo {journal} {Phys. Rev. B}}\ }%
  \textbf{\bibinfo {volume} {46}},\ \bibinfo {pages} {15233} (\bibinfo {month}
  {Dec}\ \bibinfo {year} {1992}),\
  \url{http://link.aps.org/doi/10.1103/PhysRevB.46.15233}%
  \bibAnnoteFile{NoStop}{Kane92b}%
\bibitem{Basset10}%
  \BibitemOpen
  \bibfield{author}{%
  \bibinfo {author} {\bibfnamefont{J.}~\bibnamefont{Basset}}, \bibinfo {author}
  {\bibfnamefont{H.}~\bibnamefont{Bouchiat}},\ and\ \bibinfo {author}
  {\bibfnamefont{R.}~\bibnamefont{Deblock}},\ }%
  \bibfield{journal}{%
  \Doi{10.1103/PhysRevLett.105.166801}{\bibinfo {journal} {Phys. Rev. Lett.}}\
  }%
  \textbf{\bibinfo {volume} {105}},\ \bibinfo {pages} {166801} (\bibinfo
  {month} {Oct}\ \bibinfo {year} {2010}),\
  \url{http://link.aps.org/doi/10.1103/PhysRevLett.105.166801}%
  \bibAnnoteFile{NoStop}{Basset10}%
\bibitem{Basset12}%
  \BibitemOpen
  \bibfield{author}{%
  \bibinfo {author} {\bibfnamefont{J.}~\bibnamefont{Basset}}, \bibinfo {author}
  {\bibfnamefont{A.~Y.}\ \bibnamefont{Kasumov}}, \bibinfo {author}
  {\bibfnamefont{C.~P.}\ \bibnamefont{Moca}}, \bibinfo {author}
  {\bibfnamefont{G.}~\bibnamefont{Zar\'and}}, \bibinfo {author}
  {\bibfnamefont{P.}~\bibnamefont{Simon}}, \bibinfo {author}
  {\bibfnamefont{H.}~\bibnamefont{Bouchiat}},\ and\ \bibinfo {author}
  {\bibfnamefont{R.}~\bibnamefont{Deblock}},\ }%
  \bibfield{journal}{%
  \Doi{10.1103/PhysRevLett.108.046802}{\bibinfo {journal} {Phys. Rev. Lett.}}\
  }%
  \textbf{\bibinfo {volume} {108}},\ \bibinfo {pages} {046802} (\bibinfo
  {month} {Jan}\ \bibinfo {year} {2012}),\
  \url{http://link.aps.org/doi/10.1103/PhysRevLett.108.046802}%
  \bibAnnoteFile{NoStop}{Basset12}%
\bibitem{Chamon95}%
  \BibitemOpen
  \bibfield{author}{%
  \bibinfo {author} {\bibfnamefont{C.~d.~C.}\ \bibnamefont{Chamon}}, \bibinfo
  {author} {\bibfnamefont{D.~E.}\ \bibnamefont{Freed}},\ and\ \bibinfo {author}
  {\bibfnamefont{X.~G.}\ \bibnamefont{Wen}},\ }%
  \bibfield{journal}{%
  \Doi{10.1103/PhysRevB.51.2363}{\bibinfo {journal} {Phys. Rev. B}}\ }%
  \textbf{\bibinfo {volume} {51}},\ \bibinfo {pages} {2363} (\bibinfo {month}
  {Jan}\ \bibinfo {year} {1995}),\
  \url{http://link.aps.org/doi/10.1103/PhysRevB.51.2363}%
  \bibAnnoteFile{NoStop}{Chamon95}%
\bibitem{Sassetti1}%
  \BibitemOpen
  \bibfield{author}{%
  \bibinfo {author} {\bibfnamefont{M.}~\bibnamefont{Carrega}}, \bibinfo
  {author} {\bibfnamefont{D.}~\bibnamefont{Ferraro}}, \bibinfo {author}
  {\bibfnamefont{A.}~\bibnamefont{Braggio}}, \bibinfo {author}
  {\bibfnamefont{N.}~\bibnamefont{Magnoli}},\ and\ \bibinfo {author}
  {\bibfnamefont{M.}~\bibnamefont{Sassetti}},\ }%
  \bibfield{journal}{%
  \Doi{10.1088/1367-2630/14/2/023017}{\bibinfo {journal} {New J. Phys.}}\ }%
  \textbf{\bibinfo {volume} {14}},\ \bibinfo {pages} {023017} (\bibinfo {month}
  {February}\ \bibinfo {year} {2012}),\
  \url{http://iopscience.iop.org/1367-2630/14/2/023017}%
  \bibAnnoteFile{NoStop}{Sassetti1}%
\bibitem{Keldysh}%
  \BibitemOpen
  \bibfield{author}{%
  \bibinfo {author} {\bibfnamefont{L.}~\bibnamefont{Keldysh}},\ }%
  \bibfield{journal}{%
  \bibinfo {journal} {Sov. Phys. JETP}\ }%
  \textbf{\bibinfo {volume} {20}} (\bibinfo {year} {1965})%
  \bibAnnoteFile{NoStop}{Keldysh}%
\bibitem{tztl2}%
  \BibitemOpen
  \bibfield{author}{%
  \bibinfo {author} {\bibfnamefont{C.-X.}\ \bibnamefont{Liu}}, \bibinfo
  {author} {\bibfnamefont{J.~C.}\ \bibnamefont{Budich}}, \bibinfo {author}
  {\bibfnamefont{P.}~\bibnamefont{Recher}},\ and\ \bibinfo {author}
  {\bibfnamefont{B.}~\bibnamefont{Trauzettel}},\ }%
  \bibfield{journal}{%
  \Doi{10.1103/PhysRevB.83.035407}{\bibinfo {journal} {Phys. Rev. B}}\ }%
  \textbf{\bibinfo {volume} {83}},\ \bibinfo {pages} {035407} (\bibinfo {month}
  {Jan}\ \bibinfo {year} {2011}),\
  \url{http://link.aps.org/doi/10.1103/PhysRevB.83.035407}%
  \bibAnnoteFile{NoStop}{tztl2}%
\bibitem{Sassetti2}%
  \BibitemOpen
  \bibfield{author}{%
  \bibinfo {author} {\bibfnamefont{G.}~\bibnamefont{Dolcetto}}, \bibinfo
  {author} {\bibfnamefont{S.}~\bibnamefont{Barbarino}}, \bibinfo {author}
  {\bibfnamefont{D.}~\bibnamefont{Ferraro}}, \bibinfo {author}
  {\bibfnamefont{N.}~\bibnamefont{Magnoli}},\ and\ \bibinfo {author}
  {\bibfnamefont{M.}~\bibnamefont{Sassetti}},\ }%
  \bibfield{journal}{%
  \Doi{10.1103/PhysRevB.85.195138}{\bibinfo {journal} {Phys. Rev. B}}\ }%
  \textbf{\bibinfo {volume} {85}},\ \bibinfo {pages} {195138} (\bibinfo {month}
  {May}\ \bibinfo {year} {2012}),\
  \url{http://link.aps.org/doi/10.1103/PhysRevB.85.195138}%
  \bibAnnoteFile{NoStop}{Sassetti2}%
\bibitem{Glattli}%
  \BibitemOpen
  \bibfield{author}{%
  \bibinfo {author} {\bibfnamefont{E.}~\bibnamefont{Zakka-Bajjani}}, \bibinfo
  {author} {\bibfnamefont{J.}~\bibnamefont{S\'egala}}, \bibinfo {author}
  {\bibfnamefont{F.}~\bibnamefont{Portier}}, \bibinfo {author}
  {\bibfnamefont{P.}~\bibnamefont{Roche}}, \bibinfo {author}
  {\bibfnamefont{D.~C.}\ \bibnamefont{Glattli}}, \bibinfo {author}
  {\bibfnamefont{A.}~\bibnamefont{Cavanna}},\ and\ \bibinfo {author}
  {\bibfnamefont{Y.}~\bibnamefont{Jin}},\ }%
  \bibfield{journal}{%
  \Doi{10.1103/PhysRevLett.99.236803}{\bibinfo {journal} {Phys. Rev. Lett.}}\
  }%
  \textbf{\bibinfo {volume} {99}},\ \bibinfo {pages} {236803} (\bibinfo {month}
  {Dec}\ \bibinfo {year} {2007}),\
  \url{http://link.aps.org/doi/10.1103/PhysRevLett.99.236803}%
  \bibAnnoteFile{NoStop}{Glattli}%
\bibitem{Safi07}%
  \BibitemOpen
  \bibfield{author}{%
  \bibinfo {author} {\bibfnamefont{C.}~\bibnamefont{Bena}}\ and\ \bibinfo
  {author} {\bibfnamefont{I.}~\bibnamefont{Safi}},\ }%
  \bibfield{journal}{%
  \Doi{10.1103/PhysRevB.76.125317}{\bibinfo {journal} {Phys. Rev. B}}\ }%
  \textbf{\bibinfo {volume} {76}},\ \bibinfo {pages} {125317} (\bibinfo {month}
  {Sep}\ \bibinfo {year} {2007}),\
  \url{http://link.aps.org/doi/10.1103/PhysRevB.76.125317}%
  \bibAnnoteFile{NoStop}{Safi07}%
\bibitem{tztl}%
  \BibitemOpen
  \bibfield{author}{%
  \bibinfo {author} {\bibfnamefont{F.}~\bibnamefont{Dolcini}}, \bibinfo
  {author} {\bibfnamefont{B.}~\bibnamefont{Trauzettel}}, \bibinfo {author}
  {\bibfnamefont{I.}~\bibnamefont{Safi}},\ and\ \bibinfo {author}
  {\bibfnamefont{H.}~\bibnamefont{Grabert}},\ }%
  \bibfield{journal}{%
  \Doi{10.1103/PhysRevB.71.165309}{\bibinfo {journal} {Phys. Rev. B}}\ }%
  \textbf{\bibinfo {volume} {71}},\ \bibinfo {pages} {165309} (\bibinfo {month}
  {Apr}\ \bibinfo {year} {2005}),\
  \url{http://link.aps.org/doi/10.1103/PhysRevB.71.165309}%
  \bibAnnoteFile{NoStop}{tztl}%
\bibitem{Chung}%
  \BibitemOpen
  \bibfield{author}{%
  \bibinfo {author} {\bibfnamefont{Y.-W.}\ \bibnamefont{Lee}}, \bibinfo
  {author} {\bibfnamefont{Y.~L.}\ \bibnamefont{Lee}},\ and\ \bibinfo {author}
  {\bibfnamefont{C.-H.}\ \bibnamefont{Chung}}}%
   (\bibinfo {year} {2012}),\
  \Eprint{http://arxiv.org/abs/1207.5080}{arXiv:1207.5080}%
  \bibAnnoteFile{NoStop}{Chung}%
\end{thebibliography}%

\end{document}